\documentclass[12pt]{iopart}
\usepackage{graphicx,epstopdf}

\expandafter\let\csname equation*\endcsname\relax
\expandafter\let\csname endequation*\endcsname\relax
\usepackage{amsmath}

\usepackage{textcomp}
\usepackage{mathrsfs}
\usepackage{amsfonts}
\usepackage{xcolor}
\usepackage{upgreek}

\begin{document}

\newcommand{\vett}[1]{\mathbf{#1}}
\newcommand{\uvett}[1]{\hat{\vett{#1}}}
\newcommand{\beq}{\begin{equation}}
\newcommand{\eeq}{\end{equation}}
\newcommand{\bseq}{\begin{subequations}}
\newcommand{\eseq}{\end{subequations}}
\newcommand{\barr}{\begin{eqnarray}}
\newcommand{\earr}{\end{eqnarray}}
\newcommand{\GH}{Goos-H$\ddot{\mathrm{a}}$nchen }
\newcommand{\IF}{Imbert-Fedorov }
\newcommand{\bra}[1]{\langle #1|}
\newcommand{\ket}[1]{| #1\rangle}
\newcommand{\expectation}[3]{\langle #1|#2|#3\rangle}
\newcommand{\braket}[2]{\langle #1|#2\rangle}           \newcommand{\red}[1]{\textcolor{red}{#1}}   
\newcommand{\blue}[1]{\textcolor{blue}{#1}} 

\title{Localized waves carrying orbital angular momentum in optical fibers}
\author{Paula Nuño Ruano$^{1,2}$, Charles W. Robson$^2$, and Marco Ornigotti$^2$}

\address{$^1$ Aix-Marseille University, Marseille, France}
\address{$^2$ Photonics Laboratory, Physics Unit, Tampere University, Tampere, Finland}

\begin{abstract}
We consider the effect of orbital angular momentum (OAM) on localized waves in optical fibers using theory and numerical simulations, focusing on splash pulses and focus wave modes. For splash pulses, our results show that they may carry OAM only up to a certain maximal value. We also examine how one can optically excite these OAM-carrying modes, and discuss potential applications in communications, sensing, and signal filtering.
\end{abstract}

\maketitle

\section{Introduction}

Pulsed light is a key feature of modern optics, being utilized for communications \cite{Agrawal_2}, remote sensing and imaging \cite{Weiner}, environmental monitoring, and even medicine \cite{Kartner}. Within the field of communications, one application is the use of optical pulses to transmit information through a fiber, but dispersive effects can lead to limitations on the bit rate and transmission distance \cite{Agrawal_2}. One way to manage this problem of pulse dispersion is to use the optically nonlinear properties of fibers to counteract it, leading to stable pulses known as solitons \cite{Agrawal_1}; for these nonlinear solitons, however, limitations exist in their usage due to fiber losses and timing jitter \cite{Agrawal_2}.

Due to these limitations, it is an attractive prospect to create localized waves in an optical fiber without the requirement of optical nonlinearity. In the early 1990s \cite{vengsarkar_closed-form_1992}, it was shown theoretically, using a bidirectional-decomposition approach \cite{besieris_bidirectional_1989}, that localized waves could indeed be sustained in optically linear fibers and several different source spectra were investigated in detail. Localized waves \cite{hernandez-figueroa_localized_2008}, i.e., nondiffracting solutions of the wave equation, that propagate without changing their shape in either space or time, and in particular X-waves, which can be considered as their most famous representative, have been studied in several different areas of physics, ranging from acoustics, where they have been firstly introduced \cite{refX1,refX2}, to nonlinear optics \cite{refX3}, Bose-Einstein condensates \cite{refX4}, quantum optics \cite{refX5,refX6}, and waveguide arrays \cite{refX7,refX8}, to name but a few. Recently, moreover, fully controllable exact paraxial nondiffracting waves have been demonstrated, using a novel spectral engineering technique \cite{Kondakci}.

Despite the progress described above, there has been a relative shortage of research on the potential orbital angular momentum (OAM) of localized pulses, with a few exceptions such as generalized X-waves in free space \cite{ornigotti_effect_2015,refX9}, as well as in nondispersive media \cite{refX6}. 

To the best of our knowledge, no attempt at describing the properties of localized waves carrying OAM in optical fibers has yet been made, and these kinds of fiber modes still remain unknown. In the present work, therefore, we show that localized waves in an optically linear fiber can be generalized to waves carrying OAM. We thoroughly discuss their properties and how to couple them using suitably structured light pulses, and we also notice the presence, for some of these localized waves, of a maximum value of the OAM they can carry inside the fiber. This may prove useful for future applications as the extra degrees of freedom provided by OAM are known to be able to increase the capacity for information encoding and also to improve the accuracy of light-based sensing \cite{J_Wang,Robert,Leone}.

In this work, we base our theory on the bidirectional-decomposition method \cite{vengsarkar_closed-form_1992,besieris_bidirectional_1989,Ziolkowski_aperture} mentioned above -- a decomposition of solutions to the wave equation into a product of forward- and backward-propagating plane waves -- as this is a natural approach to the study of localized waves and, as we will show, is particularly suitable for the generalization of previous results \cite{vengsarkar_closed-form_1992} relating to localized waves to those carrying OAM.

This work is organized as follows: in Section 2, we briefly review the propagation of a scalar electromagnetic field in a cylindrically symmetric, step-index optical fiber, and we show how it is possible to express a general solution of the wave equation in terms of a bidirectional decomposition. This section will mostly serve to fix the notation used throughout our manuscript. Then, in Sections 3 and 4, we present two classes of localized waves carrying OAM, namely focus wave modes (FWM) and splash pulses, respectively. For each class, we investigate their main features and discuss how it is possible to couple into such modes using twisted optical pulses. Finally, conclusions are drawn in Section 5.

\section{Scalar Waves in Optical Fibers}
\subsection{General Solution}
\begin{figure}[t!]
    \centering
    \includegraphics[width=\columnwidth]{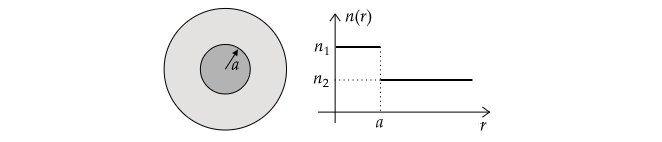}
    \caption{Schematic representation of the transverse cross section (left) and refractive index profile (right) for a cylindrically symmetric, step-index optical fiber. $a$ is the fiber radius, while $n_{1,2}$ are the core and cladding refractive indices, respectively.}
    \label{fig:fiber}
\end{figure}
The propagation of a scalar electromagnetic field in a step-index optical fiber made of a nondispersive material (Fig.~\ref{fig:fiber}) can be described by the following wave equation \cite{snyderLove}
\begin{equation} \label{eq:scalar}
    \left(\nabla - \frac{n_i^2}{c^2} \partial_t^2 \right) \psi_i(\vett{r},t) = 0,
\end{equation}
where $\nabla$ is the three-dimensional Laplacian, $n_{1,2}$ the refractive index of the fiber core and cladding, respectively (with $n_1>n_2$), and $c$ is the speed of light in vacuum. 

A general solution to the above equation in terms of localized waves can be written using the so-called bidirectional decomposition \cite{besieris_bidirectional_1989} as
\begin{equation} \label{eq:bidirectional}
    \psi(\vett{r},t) = \Phi(\rho,\varphi) e^{-i\alpha \xi} \, e^{i \beta \eta}
\end{equation}
where $\xi = z-ct$ and $\eta = z+ct$ represent the forward and backward co-moving propagation directions, respectively, upon which the bidirectional decomposition is based, and $\alpha,\beta$ are their correspondent propagation constants.

Decomposing the Laplacian in Eq.~\eqref{eq:scalar} into its longitudinal ($\partial^2_z$) and transverse ($\nabla_{\perp}^2$) components, the scalar wave equation reduces to the Helmholtz equation $(\nabla_\perp + \kappa_i^2)\Phi = 0$,
where $\kappa_i^2 = n_i^2(\alpha + \beta)^2 - (\alpha - \beta)^2$. In cylindrical coordinates, the general solution, compatible with propagation inside an optical fiber, is given in terms of Bessel functions $J_m(\kappa_1\rho)$ in the core and modified Bessel functions $K_m(\kappa_2\rho)$ in the cladding \cite{snyderLove}. The transverse wave vectors $\kappa_{1,2}$ are then related to the longitudinal ones from the bidirectional decomposition by the relations $\kappa_1^2 = 4\alpha\beta$, and $\kappa_2^2 = (\alpha-\beta)^2-n_e^2(\alpha+\beta)^2$, where $n_e=n_2/n_1$ is the effective refractive index.

With this at hand, we can therefore write the general pulse-like solution to Eq. \eqref{eq:scalar} within the framework of the bidirectional decomposition as follows
\begin{subequations}\label{eq:core}
\begin{align}
\psi_m(\vett{r},t) &= \frac{e^{im\varphi}}{(2\pi)^2}\int\,\mathrm{d}\alpha\,\mathrm{d}\beta\,\mathrm{d}\kappa_1\, g_m(\alpha,\beta,\kappa_1)\,\kappa_1 J_m(\kappa_1 \rho) \, e^{i(-\alpha\xi+\beta\eta)}\,\delta\left(\kappa_1^2-4\alpha\beta\right)\\
\psi_m(\vett{r},t) &= \frac{\,e^{im\varphi}}{(2\pi)^2}  \int \mathrm{d}\alpha\,\mathrm{d}\beta\,\mathrm{d}\kappa_1\,d\kappa_2\, g_m(\alpha,\beta,\kappa_1)\, \kappa_1\,K_m(\kappa_2 \rho)\,e^{i\,(-\alpha\xi+\beta\eta)}\nonumber\\ 
&\times\delta(\kappa_1^2-4\alpha\beta) \, \delta(\kappa_2^2 -w(\alpha,\beta) ) \, \delta\left(\kappa_2 - f_m(\kappa_1)\right),
\end{align}
\end{subequations}
where the first expression is valid within the fiber core (i.e., $\rho \leq a$), while the second in the cladding ($\rho> a$). Notice, moreover, that in the above equations we have defined $w(\alpha,\beta)=(\alpha - \beta)^2 - n_e^2(\alpha + \beta)^2$ as the wave guiding condition, and $f_m(\kappa_1)$ as the characteristic equation for the mode $m$ (see Appendix A), which implicitly implements the constraint imposed in the choice of $\kappa_2$ once $\kappa_1$ is fixed through the characteristic equation of the fiber.
\section{Focus Wave Modes Carrying OAM}
The first class of OAM-carrying localized solutions we are going to investigate is represented by the so-called Focus Wave Modes (FWMs), characterized by the following spectrum \cite{belanger, sezginer, ziolk2}
\begin{equation} \label{eq:FWMspectrum}
    g(\alpha,\beta,\kappa_1) = 8\pi^2 a_1 \, e^{-a_1\alpha} \delta(\beta - \beta_0)
\end{equation}
where $a_1$ has the dimensions of a length and it represents the spectral width of the pulse, and $\beta_0$ is an arbitrarily chosen, but constant, propagation constant in the $\eta$-direction. To find an analytical solution to Eq. \eqref{eq:core} with the spectrum above, we first choose the integration order to be $\int_0^{\infty}d\kappa_1\,\int_0^{\infty}d\beta\,\int_{\beta s}^{\infty}d\alpha$ (see Appendix A). This, in fact, will allow us to take advantage of the Dirac delta function in the spectrum above and write the solution in the core as follows:
\beq
\psi_{FWM}(\vett{r},t)  = a_1 e^{i(\beta_0 \eta + m\varphi)} \int_{\beta_0s}^\infty \mathrm{d}\alpha J_m\left(2\rho\sqrt{\alpha\beta_0}\right) e^{-\alpha(a_1+i\xi)}.
\eeq
If we now split the above integral into its unperturbed and wall components (see Appendix A), we obtain, with the help of Eq. (6.614.1) of Ref. \cite{gradshteyn_table_2007}, the following closed-form expression for the unperturbed term (see Appendix B for details on the calculation)
\barr\label{eq:FWMu_dim}
\psi_u(\vett{r},t) &=& \frac{a_1\rho\sqrt{2\pi}}{\beta_0 w^3(\xi)} e^{-\frac{\rho^2}{w^2(\xi))}+i(\beta_0 \eta + m\varphi)}F_m\left(\frac{\rho^2}{w^2(\xi)}\right),
\earr
where $F_n(x)=I_{(n-1)/2}(x)-I_{(n+1)/2}(x)$, with $I_n(x)$ being the modified Bessel function of the first kind \cite{nist}, and
\beq
w^2(\xi)=w_0^2\left(1+i\frac{\xi}{\xi_R}\right),
\eeq
where, in analogy with Gaussian beams, $w_0^2=2a_1/\beta_0$ is the transverse width of the pulse in $\xi=0$, and $\xi_R=a_1$ is the ``Rayleigh range" of the pulse. The spectral width, therefore, allows one to control the localization and width of the pulse, while the parameter $\beta_0$ essentially regulates only its initial size. From the literature, a broad bandwidth, i.e., $a_1 \ll 1$, is necessary for localization along propagation \cite{besieris_bidirectional_1989}, while simultaneously $a_1 \beta_0 \ll 1$ must hold to ensure causality \cite{hernandez-figueroa_localized_2008}.
The wall term, on the other hand, does not admit a simple analytical form and must be then evaluated numerically using the following integral form
\beq\label{eq:FWMwall_dim}
\psi_w(\vett{r},t) = -2sa_1 \beta_0 e^{i(\beta_0 \eta + m\varphi)} \int_0^1 \mathrm{d} x \, x \, J_m(\omega_0 x) \, e^{-x^2/\Delta x^2},
\eeq
where  $\omega_0 = 2\rho\beta_0\sqrt{s}$ and $\Delta x^2 = 1/s\beta_0(a_1+i\xi)$.

This is the first result of our work: the above expressions describe an FWM carrying OAM propagating along a step-index optical fiber. 
%
\begin{figure}[t!]
    \centering
    \includegraphics[width=\columnwidth]{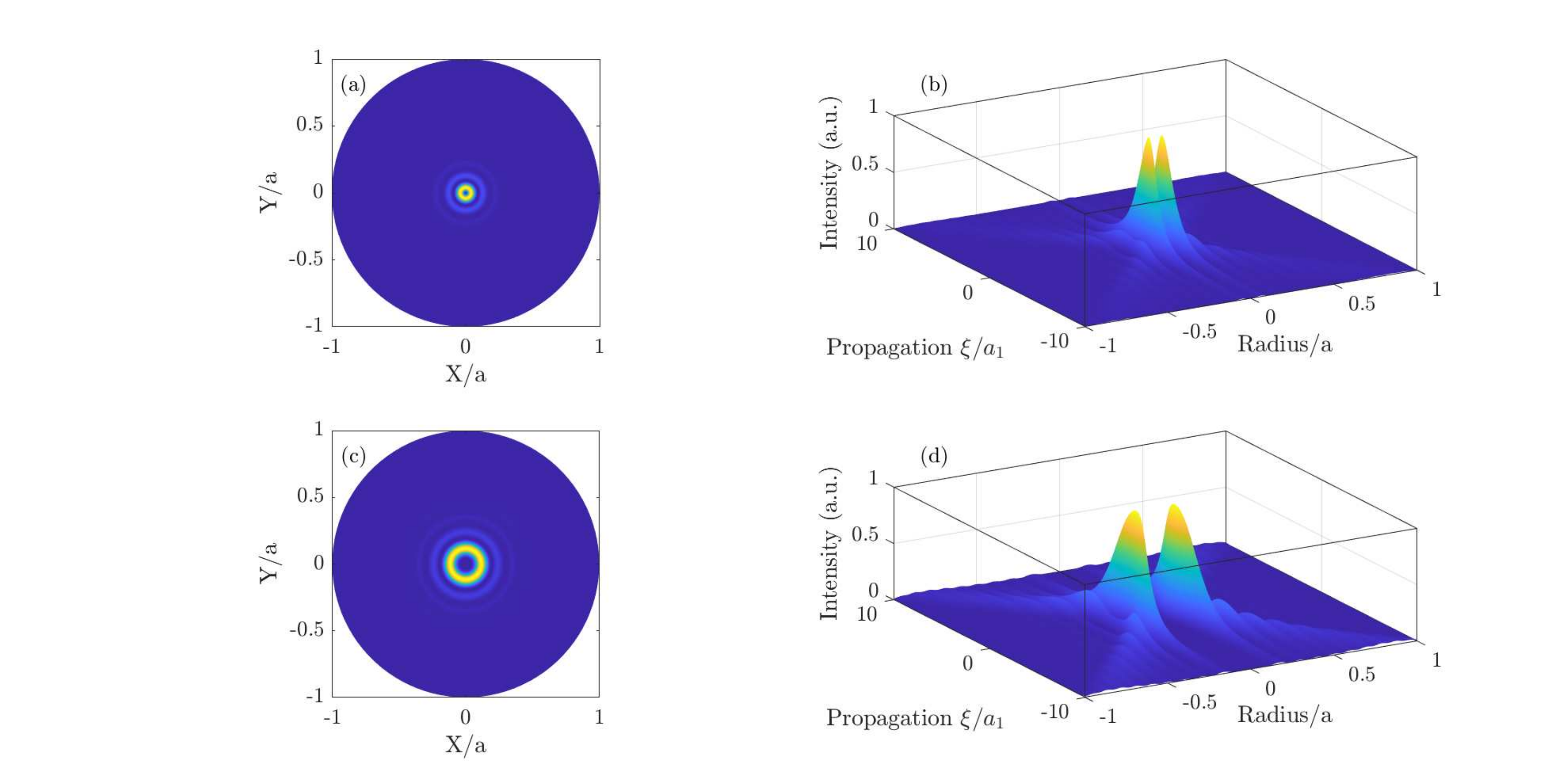}
    \caption{Evolution of an FWM pulse carrying $m = 1$ (top row) and $m = 4$ (bottom row) units of angular momentum as it propagates through the fiber. (a,c) Transverse section at $z = ct = 0$ for the normalized total solution, $\left|\psi_u + \psi_w\right|^2$. (b,d) Normalized total solution, $\left|\psi_u + \psi_w\right|^2$. The parametric values used are $a = 10\, \upmu$m, $a_1 = 2 \cdot 10^{-7}$ m, $\beta_0 = 5 \cdot 10^5 \text{ m}^{-1}$ and $s = 10$. }
    \label{fig:FWMpulse}
\end{figure}

The transverse and longitudinal structure of the complete OAM-carrying FWMs in optical fibers, i.e., including both unperturbed and wall term contributions, is depicted in Fig.~\ref{fig:FWMpulse}, for different values of the OAM parameter $m$. As can be seen, the transverse field distribution is well-confined within the fiber core and consists of several rings and a central doughnut-shape, typical of OAM-carrying modes. The longitudinal structure, on the other hand, presents an X-shape with no intensity in its central region, which mimics the zero-intensity region of its transverse profile, in a similar manner of OAM-carrying X-waves \cite{ornigotti_effect_2015}, due to the fact that both localized waves are, essentially, polychromatic Bessel beams with exponentially-decaying spectra. However, the different dependence of the argument of the Bessel function with respect to the integration variable ($\sqrt{\alpha}$ for FWMs, and $\alpha$ for X-waves - see below) makes their actual transverse profile different in nature, although they both possess the same features, i.e., ring structure and OAM. The radius of the ring in the transverse section or, equivalently, the width of the central zero-intensity line along the propagation direction increases with the angular momentum as a result of the stronger associated phase singularity. This pulse presents a high radial localization and a longitudinal extension that increases with the angular momentum.

\begin{figure}[!t]
    \centering
    \includegraphics[width=0.6\textwidth]{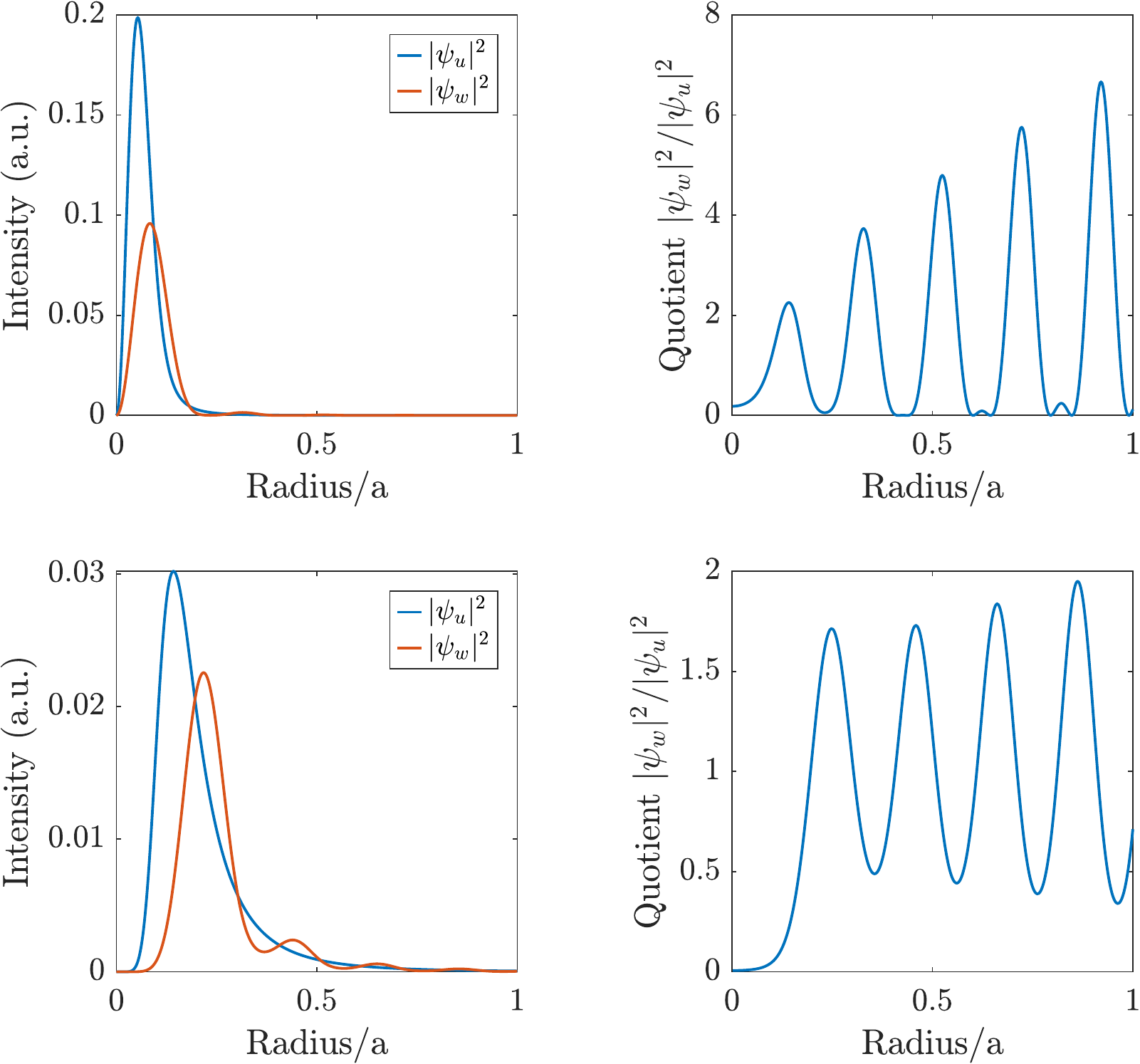}
    \caption{Comparison between the unperturbed (blue) and the wall (red) term at $z = ct$ for an FWM pulse, and their relative quotient $\left|\psi_w/\psi_u\right|^2$ for different values of the OAM parameter $m=1$ (top row), and $m=4$ (bottom row). The parameters take the values $a = 10$ $\mu$m, $a_1 = 2 \cdot 10^{-7}$ m, $\beta_0 = 5 \cdot 10^5 \text{ m}^{-1}$ and $s = 10$.}
    \label{fig:FWM_comparison}
\end{figure}

To understand the relative weight of the unperturbed and wall term in determining the final form of the localized pulse, in Fig. \ref{fig:FWM_comparison} we plot the ratio $|\psi_w/\psi_u|^2$, for the same values of OAM used in Fig. \ref{fig:FWMpulse}, namely $m=1$, and $m=4$. As can be seen, the radial profiles of $\psi_{u,w}$ are very similar, and overlap quite well radially. As $m$ increases, they overlap less (a feature that is due to the different functional dependencies of the two fields on the OAM parameter $m$) and ripples start to appear in the wall term, close to the tail of $\psi_u$. From Fig. \ref{fig:FWM_comparison}, we can also see how the wall term $\psi_w$ has a richer radial structure than $\psi_u$ and, de facto, cannot be interpreted as a small perturbation of $\psi_u$ since its weight gets progressively bigger as the field approaches the core/cladding edge. This means simply that $\psi_w$ contains information on the correct boundary conditions (i.e., the presence of the core/cladding edge) that need to be applied to $\psi_u$ to adapt it to the optical fiber itself.
\subsection{Coupling of OAM-Carrying FWM with an X-Wave}
\begin{figure}[t!]
    \centering
    \includegraphics[width=\columnwidth]{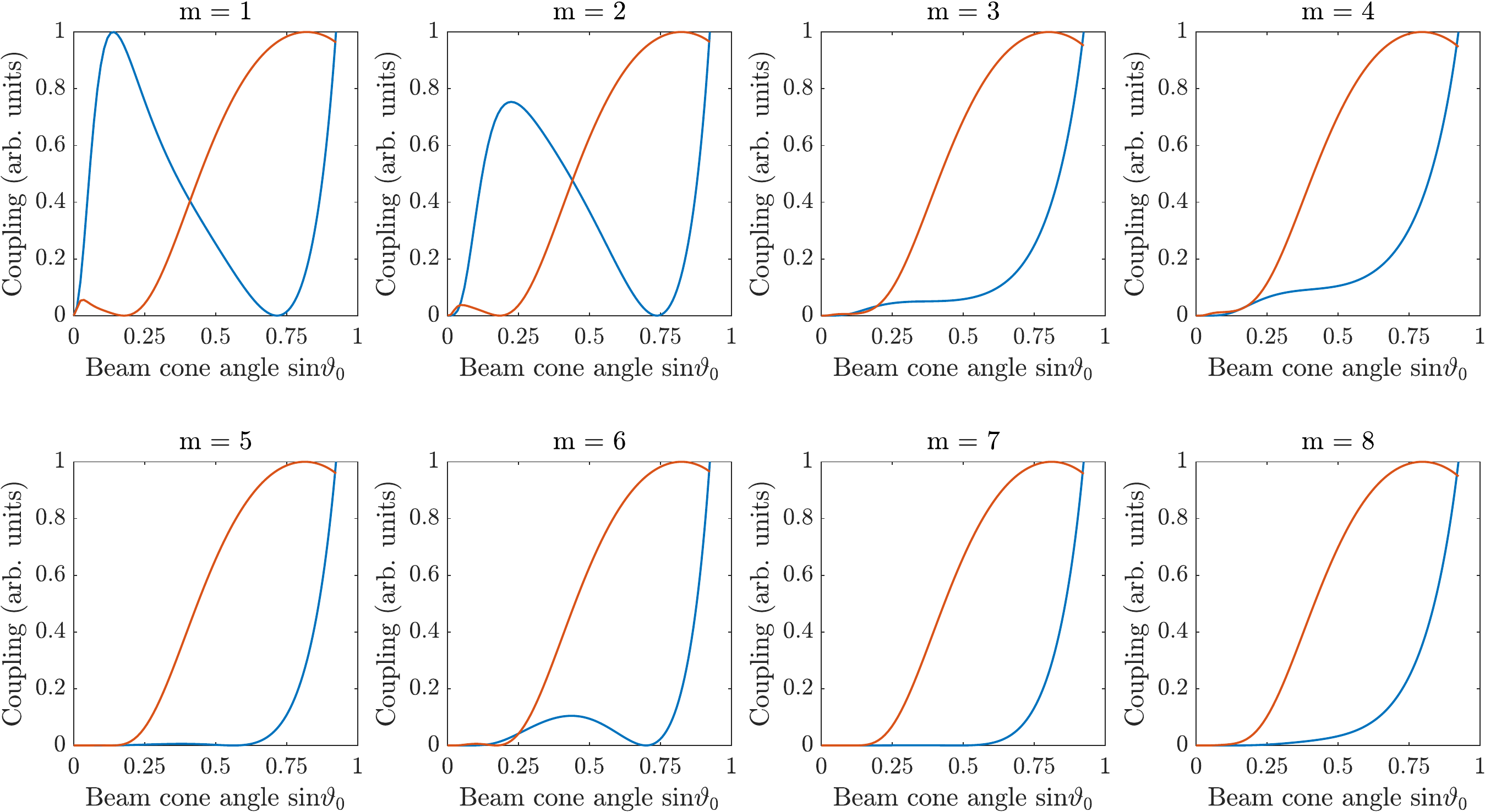}
    \caption{Coupling between an FWM pulse and an incoming X-wave, both having the same radial scale $c\alpha = \sqrt{2a_1/\beta_0}$ (blue line) or the same bandwidth $c\alpha = a_1$ (red line). The parameters used for these plots are $a = 10\, \upmu$m, $a_1 = 2 \cdot 10^{-7}$ m, $\beta_0 = 5 \cdot 10^5 \text{ m}^{-1}$ and $s = 10$.}
    \label{fig:FWM_Xwave_coupling}
\end{figure}
The existence of localized waves carrying OAM in optical fibers is quite interesting, as they are naturally resilient to external perturbations and losses, and could be employed as a solid mean of communication using OAM in fibers. A crucial question is then, how is it possible to excite such electromagnetic field configurations in an optical fiber, when coupling light into it. The easiest solution to this problem would be to couple a localized wave carrying OAM in the fiber, in order to excite the correspondent fiber mode. X-waves, in particular, are a viable solution for this task, given the similar structure they possess with respect to FWMs. 

Although a thorough study would require knowledge of the form of the mode in both the core and cladding region, in this section we limit our analysis to the study of the coupling in the core region solely. This approximation is justified within the limit of the paraxial approximation (absence of strong focusing for the coupling wave) and by the fact that, as can be seen from Fig. \ref{fig:FWM_comparison}, the FWM is well-confined within the core region. 

We then consider a fundamental X-wave carrying $m$ units of OAM impinging upon the input facet of an optical fiber. According to Ref.  \cite{ornigotti_effect_2015}, its explicit expression is given by
\beq\label{eq:Xwave_dim}
    X_m(\textbf{r},t) = e^{im\varphi} \, \frac{\zeta^m}{2^m(\alpha+i\xi)} \, {_2F_1}\left(\frac{m+1}{2},\frac{m+2}{2}; m+1;-\zeta^2\right)
\eeq
where $\zeta = (\rho \sin\vartheta_0)/[c\alpha(1+i\xi)]$ and $\xi = (z/c)\cos\vartheta_0 - t$. Notice that because of its exponentially-decaying spectrum, the field above carries infinite energy and it therefore cannot be physically realized. Experimentally feasible OAM-carrying X-waves, i.e., the so-called Bessel-X pulses \cite{Bessel_X_pulse_book}, on the other hand, can be generated using a Gaussian rather than an exponentially-decaying spectrum. Apart from carrying finite energy, Bessel-X pulses retain all the salient features of fundamental X-waves, up to a certain spatial/temporal scale, determined by their Gaussian spectrum. However, they are harder to manipulate analytically. For this reason, for the rest of the manuscript we will employ fundamental X-waves, since they are easier to handle analytically. We will, however, point out, where the use of Bessel-X pulses could solve the infinite energy issue.

To estimate how well we can excite an FWM in the fiber using an X-wave with OAM, we calculate the overlap integral (coupling coefficient) between these two fields at the input facet of the fiber, i.e., at $z=ct=0$ (corresponding to $\eta=\xi=0$), as follows:
\begin{equation}\label{couplingC}
    C_m =\braket{X_m(r,t)}{\psi_{FWM}(r,t)}=\frac{1}{E_X} \, \left|\int_0^{2\pi}\,d\varphi\,\int_0^{\infty}\, d\rho\,\rho\,\psi_\ell^*(\rho,\varphi) \, X_m(\rho,\varphi)\right|^2
\end{equation}
where $E_X=\braket{X_m(r,t)}{X_m(r,t)}\rightarrow\infty$ is the energy of the incoming X-wave, and it therefore cannot be used to normalize the coupling coefficient. Instead, we choose to normalize $C_{l,m}$ with respect to its maximum. This is a consequence of having chosen fundamental X-waves, rather than Bessel-X pulses to work with. Notice, moreover, that since the angular dependence of both the FWM and the X-wave only appear in terms of the phase factor $\exp(im\varphi)$, it is trivial to see that the above integral forces $\ell=m$. This means, that in order to achieve efficient coupling, one must first match the OAM value of the FWM they want to excite. 

Numerical evaluations of the coupling coefficient as defined in Eq. \eqref{couplingC} are reported in Fig. \ref{fig:FWM_Xwave_coupling}, for different values of the OAM parameter $m$. As can be seen from Fig. \ref{fig:FWM_Xwave_coupling}, to calculate $C_m$ we have followed two different strategies, corresponding to the different ways to tune the parameters of the incoming X-wave to adapt to the FWM we want to excite. First, we consider an X-wave with a similar radial scale to the FWM pulse (blue line in Fig. \ref{fig:FWM_Xwave_coupling}). Then, we consider an X-wave with similar bandwidth to the FWM pulse (red line in Fig. \ref{fig:FWM_Xwave_coupling}). In the former case, the coupling increases drastically for large Bessel cone angles $\vartheta_0$ (i.e., strong focusing). Here, the scalar description used for the X-wave [i.e., Eq.~\eqref{eq:Xwave_dim}] is no longer valid, and corrections due to the vector nature of the electromagnetic field must be taken into account to obtain the correct result. Nevertheless, a relevant local maximum appears for small OAM content at small values of the Bessel cone angle $\vartheta_0$, implying that it is possible, even in the scalar and paraxial case, to achieve efficient coupling. When the incoming X-wave has a similar bandwidth to the FWM pulse, on the other hand, the coupling exhibits a similar monotonic behaviour for all values of OAM, and the results in Fig.  \ref{fig:FWM_Xwave_coupling} seem to point out the necessity of a fully vectorial analysis of the coupling, in order to correctly get an estimation of the coupling efficiency, since the maximum for the coupling occurs at large Bessel cone angles $\vartheta_0$.

This is the second result of our work. To efficiently excite an OAM-carrying FWM in an optical fiber using an impinging X-wave, the more favourable scenario, within the scalar and paraxial approximations, is to match their radial scales (i.e., $c\alpha/\sin\vartheta_0=\sqrt{2a_1/\beta}$). In this case, it is always possible to find, for small values of $m$, a paraxial value of the Bessel cone angle characterizing the X-wave, such that the coupling is maximized. For higher values of $m$, however, this optimal angle disappears and the bandwidth-matching method always offers a more efficient coupling than the waist-matching scenario.

\section{Splash Modes Carrying OAM.}
\begin{figure}[t!]
    \centering
    \includegraphics[width=\columnwidth]{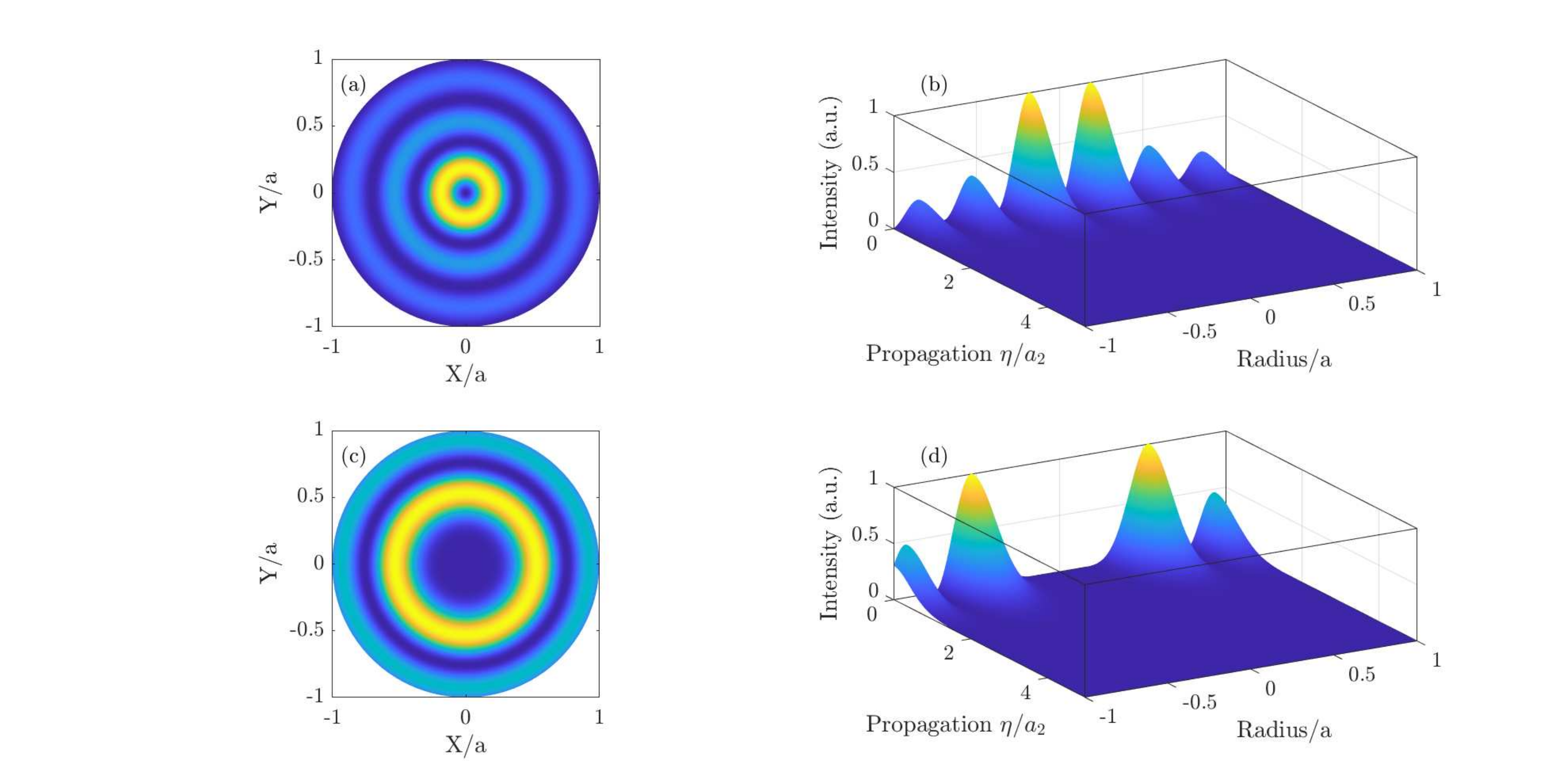}
    \caption{Evolution of a splash pulse carrying $m = 1$ (top row) and $m = 4$ (bottom row) units of angular momentum as it propagates through the fiber. (a,c) Transverse section at $z = ct = 0$ for the normalized total solution, $\left|\psi_u + \psi_w\right|^2$ . (b,d) Normalized total solution, $\left|\psi_u + \psi_w\right|^2$ for $z = ct$. The parametric values used are $a = 10 \, \upmu$m, $\gamma = 10^6 \text{ m}^ {-1}$, $a_1 = a_2 = 10^{-5}$ m and $s = 10$.}
    \label{fig:Splash_pulse}
\end{figure}
A second class of localized waves in optical fibers carrying OAM that have interesting properties are the so-called splash pulses, which are obtained by substituting the following spectrum \cite{vengsarkar_closed-form_1992} in Eqs. \eqref{eq:core}
\begin{equation} \label{eq:splashspectrum}
    g(\alpha,\beta,\kappa_1) = 8 \pi^2 \, a_1 \, a_2 \, \delta(\kappa_1 - \gamma) \, \exp\left[-(a_1\alpha + a_2\beta)\right].
\end{equation}
As in the case of FWMs, while the unperturbed term admits the following closed-form solution (see Appendix B for details on the derivation)
\begin{equation} \label{eq:splash_u_dim}
    \psi_u = a_1 a_2 \gamma \, e^{im\varphi} \, J_m(\gamma \rho) \, K_0\left[\gamma \sqrt{(a_1+i\xi)(a_2-i\eta)}\right] 
\end{equation}
the wall term has no general closed-form solution, but it can be represented by the following integral
\begin{equation} \label{eq:splash_w}
    \psi_w = - a_1 a_2 \gamma \, e^{im\varphi} J_m(\gamma\rho) \int_0^1 \frac{dx}{x} \exp\left[-\frac{\gamma}{2}\sqrt{s}(a_1+i\xi)x\right] \exp\left[-\frac{\gamma}{2x\sqrt{s}}(a_2-i\eta)\right].
\end{equation}
Notice that both the unperturbed and wall term of a splash pulse, as defined above, are separable functions of the transverse and longitudinal coordinates, i.e., they can both be written as $\psi_{u,w}(\rho,\varphi,\zeta)=f(\rho,\varphi)g(\zeta)$. Moreover, since the transverse part of both pulses is the same, i.e., $J_m(\gamma\rho)\exp{(im\varphi)}$, even the sum of these two functions, namely the overall pulse, is a separable function of transverse and longitudinal coordinates, i.e., $\psi(\rho,\varphi,\zeta)=\psi_u(\rho,\varphi,\zeta)+\psi_w(\rho,\varphi\zeta)=F(\rho,\varphi)G(\zeta)$. The only parameter linking the transverse and longitudinal shape of the splash pulse is $\gamma$, i.e., the transverse size of the pulse itself. This means, in principle, that the transverse and longitudinal shape of splash pulses carrying OAM in an optical fiber can be controlled and modified independently, at least at the first order in $\gamma$.

\begin{figure}[t!]
    \centering
    \includegraphics[width=0.7\columnwidth]{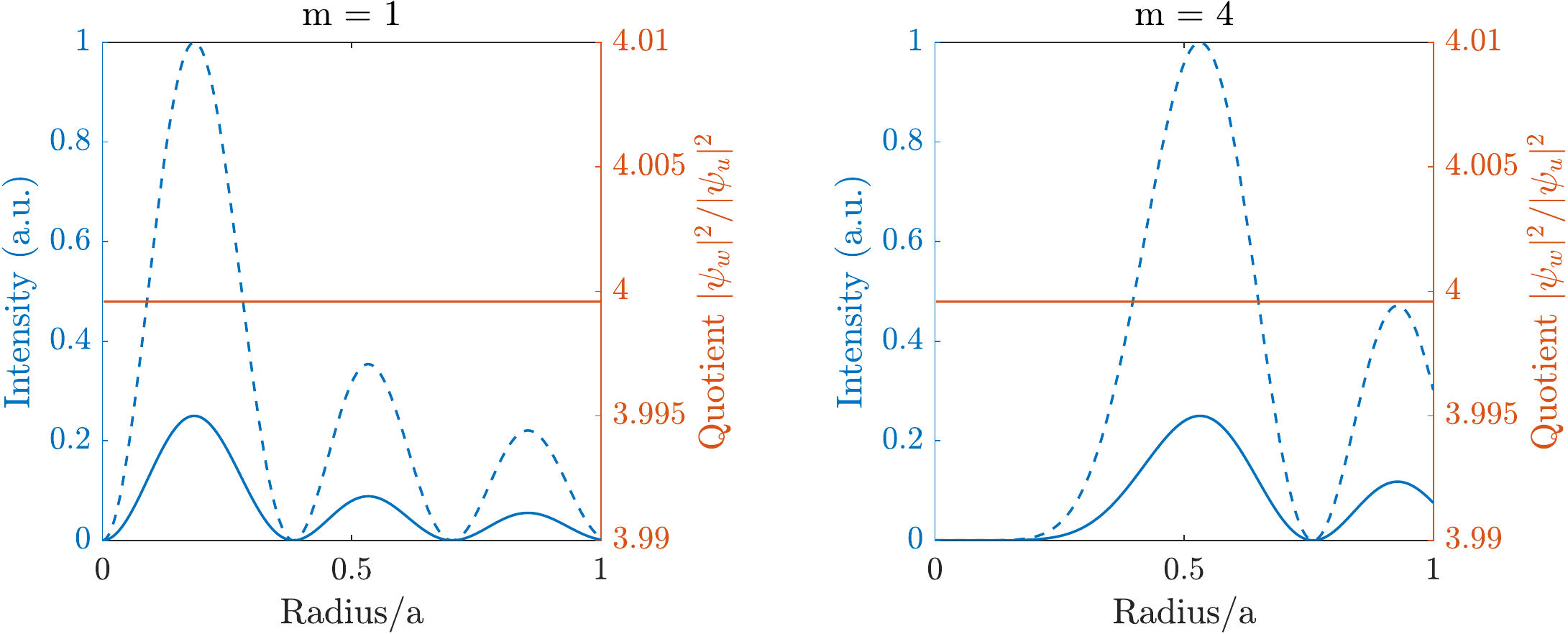}
    \caption{Comparison between the unperturbed and wall terms for a splash pulse at $z = ct = 0$ and $\varphi = 0$. On the left axis (blue), the intensity distribution of unperturbed (solid) and wall (dashed) terms. On the right axis (orange), the quotient between both intensities $\left|\psi_w\right|^2/\left|\psi_u\right|^2$. In this plot we use the values: $a = 10 \, \upmu$m, $\gamma = 10^6 \text{ m}^ {-1}$, $a_1 = a_2 = 10^{-5}$ m and $s = 10$.}
    \label{fig:splash_comparison}
\end{figure}

The longitudinal and transverse structure of splash pulses, for different values of the OAM index $m$, is reported in Fig. \ref{fig:Splash_pulse}, while a comparison between the radial structure of the unperturbed and wall terms is given in Fig. \ref{fig:splash_comparison}. As can be seen, they have the same structure, and differ only by their amplitude. Notice, that contrary to the case of FWM, the wall term of splash pulses has higher intensity (with respect to the unperturbed term) even in the very vicinity of the fiber core centre. This, once more, suggests, how the wall term cannot be considered a small perturbation, but yet an extra term to include in the analysis, to properly account for the core/cladding boundary conditions, which reflect themselves on the overall structure of the fiber mode, and not only act of its edges.
%

\begin{figure}[t!]
    \centering
    \includegraphics[width=0.5\columnwidth]{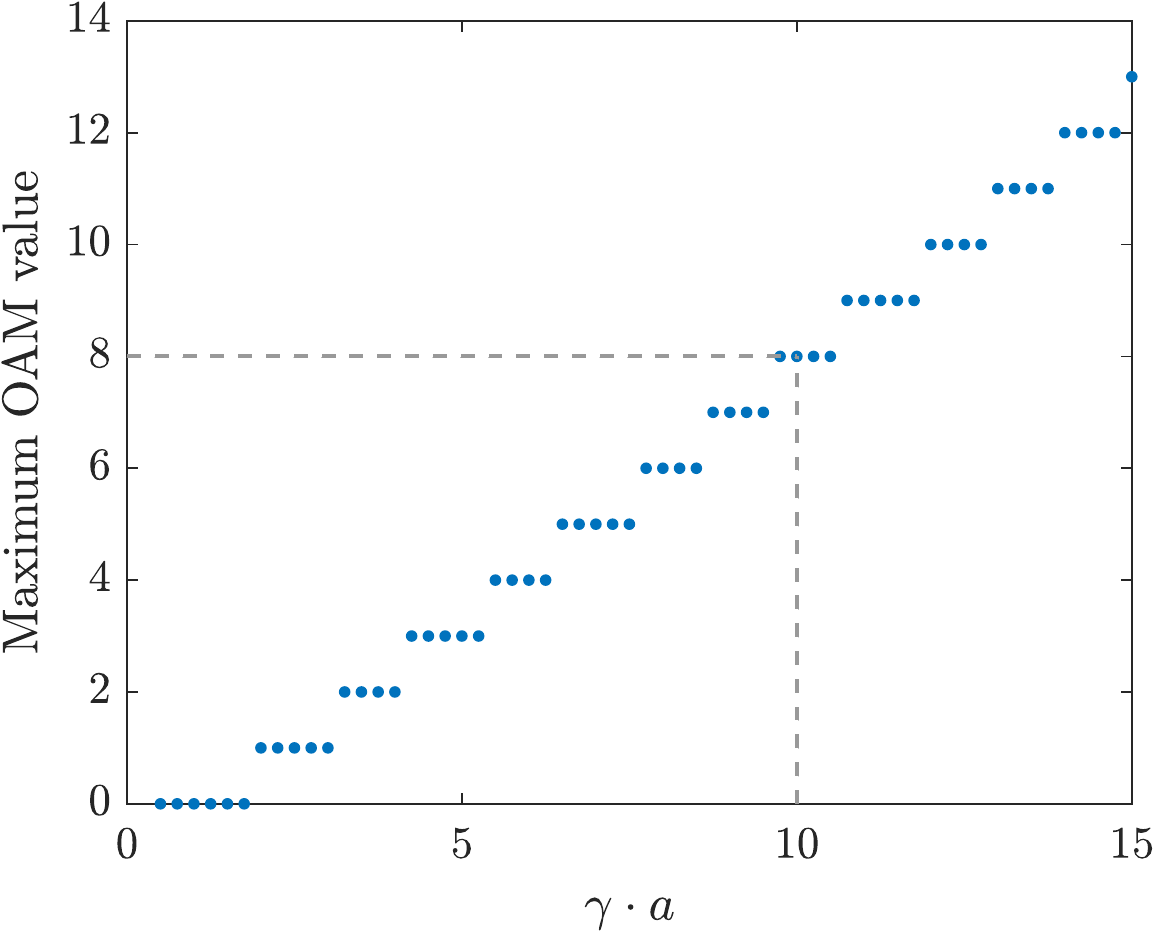}
    \caption{Maximum allowed value of angular momentum for the splash pulse to be localized inside the core. The dashed line marks the situation corresponding to  $a = 10 \, \upmu$m and $\gamma = 10^6 \text{ m}^ {-1}$ in the text, which leads to a maximum value $M=8$.}
    \label{fig:maxOAM}
\end{figure}

From Fig. \ref{fig:Splash_pulse}, and from the explicit form of the radial function of the splash pulse, we can also notice that as the OAM parameter $m$ increases, the transverse size of the doughnut also increases to compensate the higher order phase singularity carried by the pulse itself. For a splash pulse in the core region of an optical fiber, this means that there exists a maximum value of OAM such a pulse can carry, since once the doughnut ring reaches the core/cladding boundary, it must match with an exponentially-decaying-like function (at least in intensity), and cannot therefore grow any further. Higher values of OAM then result in OAM-modes not being guided anymore as a core mode of the fiber. To quantify this limit on the OAM carried by splash pulses, we can take the situation in which the ring-shaped mode has a radius $\rho=a$, i.e., when the radius of the doughnut matches exactly the size of the fiber core. The value of $m$ corresponding to this case will then represent the maximum value $m=M$ of OAM that the splash pulse can carry. This value can be calculated by recalling that for $\rho=a$,  $J'_M(\gamma a) = 0$ and $J''_M(\gamma a) < 0$ must hold (the prime indicates the derivative with respect to the radial coordinate). As it can be seen from Fig. \ref{fig:maxOAM}, these two conditiona lead to a linear relation between the maximum OAM carried by the splash pulse (M), and the scaled core radius of the fiber $\gamma a$. For example, if we use $a = 10\, \upmu$m and $\gamma = 10^6 \text{ m}^{-1}$, we find, from Fig. \ref{fig:maxOAM}, that the splash pulse can carry a maximum of $M = 8$ units of OAM.

This is the third result of our work. Splash pulses in optical fibers can only carry a limited amount of OAM. The limit to this is essentially given by the radius of the core of the fiber, and corresponds to the situation in which the radial, doughnut-shaped part of the splash pulse has a transverse size of the order of the fiber core itself. 
\subsection{Coupling of OAM-Carrying Splash Modes with Laguerre-Gaussian Beams}
The transverse section of splash pulses carrying OAM, formally a Bessel function of the first kind of order $m$,  suggests that it could be possible to excite these modes with an incoming free-space pulse, whose transverse structure is that of a Bessel-Gauss (BG) beam. However, since in the paraxial regione BG beams can be well-approximated with  Laguerre-Gaussian (LG) beams, we use the latter for the sake of simplicity \cite{saleh_fundamentals_2007}.

The coupling coefficient, defined for $z=ct=0$, will be given by
\begin{equation}
    C_{m,\ell}=\braket{LG_p^{\ell}(r,\varphi)}{\psi_{splash}}= \frac{1}{E_{LG}} \, \left|\int_0^\infty\, d\rho\,\rho \int_0^{2\pi}\,d\varphi\, \psi_m^*(\rho,\varphi) \, \mathrm{LG}_p^\ell(\rho,\varphi)\right|^2
\end{equation}
where $E_{LG}=\braket{LG_p^{\ell}(r,\varphi)}{LG_p^{\ell}(r,\varphi)}$ is the energy of the incoming LG beam. As in the case of coupling of FWMs with X-waves, we only consider coupling with the core part of the overall fiber mode. Moreover, azimuthal integration dictates that the OAM of the incoming pulse must match the OAM of the particular splash mode that we want to excite, so the condition $\ell=m$ will be implicitly assumed throughout the rest of this section. 

\begin{figure}[t!]
    \centering
    \includegraphics[width=\columnwidth]{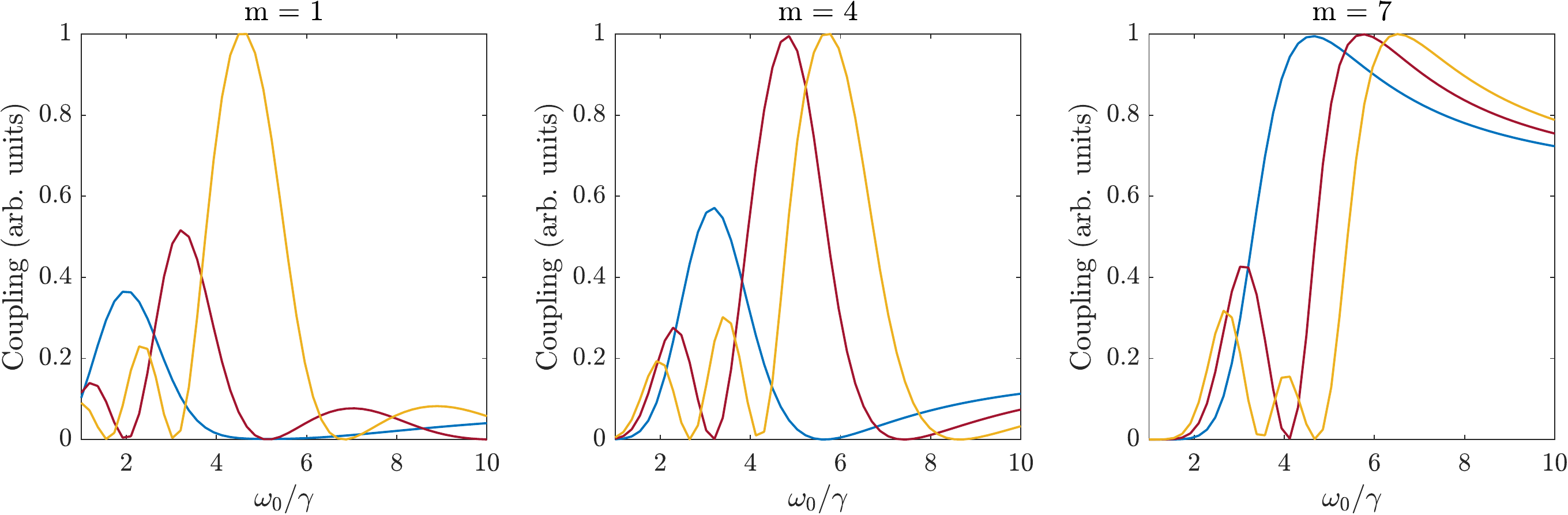}
    \caption{Coupling coefficient between an incoming LG beam and a splash pulse as a function of the normalized radial scale $w_0/\gamma$, where $w_0$ is the beam waist of the LG beam, and $\gamma$ accounts for the transverse size of the splash pulse. The plots are done for different values of $m$ and for the case of $p=0$ (solid blue line), $p=1$ (solid red line), and $p=2$ (solid yellow line).}
    \label{fig:Splash_LG_coupling}
\end{figure}

\begin{figure}[t!]
    \centering
    \includegraphics[width=\columnwidth]{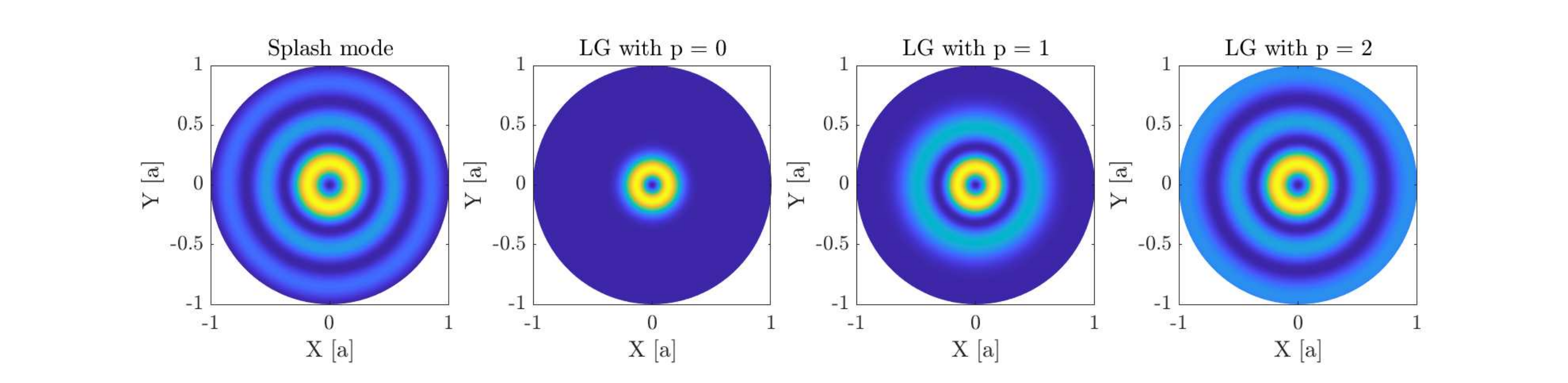}
    \caption{Comparison between the transverse section of the splash pulse (left) and the transverse structure of a LG beam with the radial scale corresponding to the maximum coupling for $p=0$, $p=1$, and $p=2$, respectively. In all these plots, $m=1$ has been used.}
    \label{fig:Splash_LG_m1}
\end{figure}

As can be seen from Fig. \ref{fig:Splash_LG_coupling}, the coupling presents a maximum value when there is a correspondence between the radial structure of the splash pulse and that of the LG beam. The closer the difference between the position of the two radial maxima is, the greater would be the coupling. To illustrate this behaviour, let us consider the example of $m=1$ (Fig.~\ref{fig:Splash_LG_m1}). When the incoming LG beam consists of a single ring, the maximum coupling is achieved on a radial scale in which this ring corresponds to the first ring in the radial structure of the splash pulse. If we consider now an LG beam with two rings, i.e., we let $p=1$, the maximum coupling is now achieved by scaling the impinging LG beam in such a way that its transverse structure now matches the first two rings of the splash pulse. The value of the coupling coefficient in this case is greater than for a single-ringed LG beam. Finally, when the LG beam structure consists of three rings, i.e., $p=2$, the maximum coupling is obtained when the scaling of the incoming LG beam is such that its transverse structure matches the transverse shape of the splash pulse. In this case, moreover, since the overlap between the two transverse modes is the maximum achievable, the coupling coefficient will also be maximum.
\section{Conclusions}
In this work, we have extended the analysis pursued by Vengsarkar and co-workers in Ref. \cite{vengsarkar_closed-form_1992} on localized waves in optical fibers, by explicitly considering the effect of OAM on them. In particular, we have focused our attention on focus wave modes and splash pulses, as they represent the most interesting (and experimentally realisable) localized waves with OAM. For each of them, we have investigated the consequence of OAM on their structure and found, for the case of splash pulses, that there exists a maximum value of OAM they can carry, which ultimately depends on the size of the core of the fiber. For each of the two classes of OAM-carrying localized waves, we have discussed how it would be possible to excite them using suitably shaped optical pulses. 

Our work shows how localized waves with OAM could represent a viable resource for fiber optics communication systems, since they combine their natural resilience against external perturbation, which guarantees their distorsion- and loss-free propagation through the fiber for long lengths, with the dense coding opportunity offered by OAM. Moreover, the limit on the maximum value of OAM carried by splash pulses can be used to engineer optical fibers, thus adapting them to the specific situation at-hand. For example, besides communication, splash-pulses-sustaining optical fibers could be used for sensing, or to filter specific values of OAM from other signals.

\section*{Acknowledgements}
The work is part of the Academy of Finland Flagship Programme, Photonics Research and Innovation (PREIN), decision 320165. P.N.R. acknowledges the support of the Erasmus Mundus grant: Erasmus+ Erasmus Mundus Europhotonics Master program (599098-EPP-1-2018-1-FR-EPPKA1-JMD-MOB) of the European Union.
\section*{Appendix A: Classical waveguide analysis and bidirectional decomposition.} \label{sec:classical vs bidirectional}
The waveguiding condition for the mode propagation constant $\beta_{cl}$ is well-known in classical analysis. In order to determine the associated constraints for $\alpha$ and $\beta$, we first need to establish a one-to-one correspondence between the bidirectional decomposition and the usual approach. In classical analysis, a possible solution is given in the form 
\begin{equation} \label{eq:classical}
    \psi(\rho,\varphi,z,t) = \Phi(\rho,\varphi) \exp(-i\beta_{cl}z) \exp(i\omega t)
\end{equation}
where $\beta_{cl}$ is the propagation constant and $\omega = 2\pi c/\lambda$ is the angular frequency. Comparing Eqs.~\eqref{eq:bidirectional}~and~\eqref{eq:classical}, we have the following correspondence
\begin{equation} \label{eq:correspondence}
    \beta_{cl} = \alpha - \beta \quad \text{and} \quad \omega = c(\alpha + \beta).
\end{equation}
The classical waveguiding condition $n_2 \omega < c \beta_{cl} < n_1 \omega$ is written in bidirectional decomposition as
\begin{equation} \label{eq:waveguiding}
    n_2 (\alpha + \beta) < \alpha - \beta < n_1(\alpha + \beta),
\end{equation}
thereby implying that $\kappa_1^2 > 0$ and $\kappa_2^2 < 0$. The explicit expression for $\kappa_{1,2}$ can be found by applying suitable boundary conditions, i.e., continuity of the field and its derivative at the core-cladding interface, which results in the usual characteristic equation for optical fibers, i.e., 
\begin{equation} \label{eq:dispersion}
    \kappa_1 \frac{J_{m+1}(\kappa_1 a)}{J_m(\kappa_1 a)} = \kappa_2 \frac{K_{m+1}(\kappa_2 a)}{K_m(\kappa_2 a)}.
\end{equation}

\begin{figure}[t!]
    \centering
    \includegraphics[width=\columnwidth]{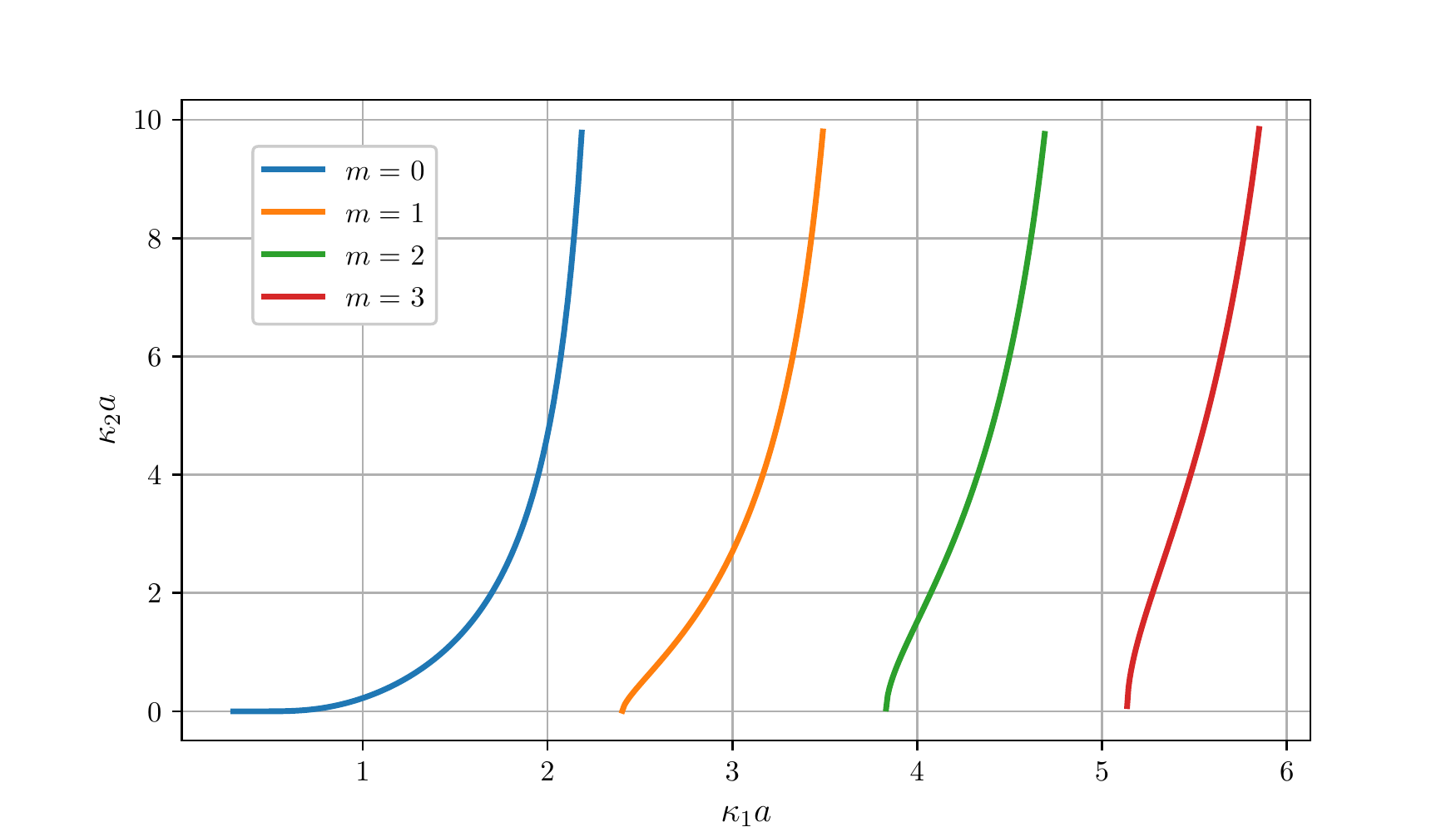}
    \caption{Functional dependency between $\kappa_2$ and $\kappa_1$ for the first modes, expressed by $f_m(\kappa_1)$. This function is defined in the text as the characteristic equation for the optical fiber. }
    \label{fig:dispersion}
\end{figure}

The numerical relation between $\kappa_1$ and $\kappa_2$ is shown in Fig. \ref{fig:dispersion}
\subsection*{Choice of the Integration Order}
Depending on the choice of the integration order in Eqs. \eqref{eq:core} above, the limits of integration for $\alpha$ and $\beta$ might vary, leading to different forms of localised solutions in the fiber. If one, for example, decides first to carry out the $\alpha$-integration, and then the $\beta$ one, the correct integration limits to be taken into account are $\int_0^{\infty}d\beta\,\int_{s\beta}^{\infty}d\alpha$, where $s=(1+n_e)/(1-n_e)$. Conversely, if the $\beta$-integration is performed first, then the integration limits become $\int_0^{\infty}d\alpha\,\int_0^{\alpha/s}d\beta$. 
It is important to notice that different choices of integration will lead to different expressions of localized waves, as is described in detail in Ref. \cite{vengsarkar_closed-form_1992}. For each of the spectra considered in this work, we will choose the appropriate order of integration that will allow us to obtain a simple, closed-form solution, which we can then handle analytically. Moreover, in the rest of this manuscript, we will limit ourselves to only investigating the form of the solution in the core of the fiber [Eq.~\eqref{eq:core}].

The expressions for the solution in the cladding region are very complicated to treat analytically, and can only be obtained numerically from the second of Eqs.~\eqref{eq:core}. As an alternative, one could empirically define an exponentially-decaying mode in the cladding and link it with the solution in the core through boundary conditions to obtain a fully analytical set of solutions in the core and cladding. However, the properties of the solution in the cladding do not influence the salient features of the localized waves in the core, and only become essential when coupling the modes of multiple optical fibers. For this reason, we then only concentrate on the core solution.
\subsection*{Unperturbed and Wall Terms}
Different choices of the modulation spectrum $g(\alpha,\beta,\kappa_1)$ in Eqs.~\eqref{eq:core} will lead, as shown in the following sections, to different localized waves. The final form of this solution, as well as the complexity of the calculations needed to obtain it, significantly depend on the integration order. As a consequence of that, and of the particular forms of the spectra we will choose, we will frequently have to deal with integral expressions of the form $\int_0^xd\xi\,\psi(\xi)$, or $\int_x^{\infty}d\xi\,\psi(\xi)$, with $\psi(\xi)$ being in general a complicated-enough function, such that the definite integral does not admit any closed-form solution. To overcome this problem, and actually deal with analytically-treatable quantities, we adopt the same strategy employed in Ref. \cite{vengsarkar_closed-form_1992}, namely we break those integrals into two terms, one referring to the unperturbed, free-space, solution, and a wall term, which accounts for the extra constraints represented by the presence of the fiber.This means, for example, that $\int_0^xd\xi\,\psi(\xi)$ could be rewritten as
\beq
\int_0^xd\xi\,\psi(\xi) = \int_0^{\infty}d\xi\,\psi(\xi) - \int_x^{\infty}d\xi\,\psi(\xi) \equiv \psi_u +\psi_w,
\eeq
where $\psi_u$ denotes the unperturbed solution, i.e., the localized wave that would exist in free space, without any fiber, and $\psi_w$ represents the so-called wall term \cite{vengsarkar_closed-form_1992}, which accounts for the corrections due to the presence of the optical fiber. The wall term then ensures that the overall solution fulfills the required boundary conditions, imposed by the fiber itself, while maintaining the overall features of the unperturbed solution. With a similar procedure, we can also introduce unperturbed and wall terms for integrals of the form $\int_x^{\infty}d\xi\,\psi(\xi)$.

\section*{Appendix B: Derivation of localized waves carrying OAM.} \label{sec:calculations}
\subsection*{Focus wave modes}
We introduce the singular spectrum [Eq.~\eqref{eq:FWMspectrum}] into the expression for the core [Eq.~\eqref{eq:core}]. Integrating first over $\kappa_1$ and using the sifting property of the Dirac delta function, we find the unperturbed $ \psi_u$ and wall $\psi_w$ terms to be
\begin{subequations}
\begin{align}
    \psi_u(\vett{r},t) & = a_1 e^{i(\beta_0 \eta + m\varphi)} \int_0^\infty \mathrm{d}\alpha J_m\left(2\rho\sqrt{\alpha\beta_0}\right) e^{-\alpha(a_1+i\xi)}, \\
    \psi_w(\vett{r},t) & = -a_1 e^{i(\beta_0 \eta + m\varphi)} \int_0^{s\beta_0} \mathrm{d}\alpha J_m\left(2\rho\sqrt{\alpha\beta_0}\right) e^{-\alpha(a_1+i\xi)}. \label{eq:wall_ap}
\end{align}
\end{subequations}

We focus first on the unperturbed term. We can obtain a closed solution by using identity~(6.614.1) from Gradshteyn and Ryzhik \cite{gradshteyn_table_2007}:
\barr
    \psi_u(\vett{r},t) & = &  \sqrt{\frac{\pi}{2}} e^{i(\beta_0 \eta + m\varphi)} \frac{a_1}{(a_1+i\xi)} \sqrt{\frac{\beta_0 r^2}{2(a_1 + i\xi)}} \exp\left(-\frac{\beta_0\rho^2}{2(a_1+i\xi)}\right)\nonumber \\ &\times& \left[I_{\frac{m-1}{2}}\left(\frac{\beta_0\rho^2}{2(a_1+i\xi)}\right) - I_{\frac{m+1}{2}}\left(\frac{\beta_0\rho^2}{2(a_1+i\xi)}\right)\right].
\earr

As for the wall term, no closed-form solution can be found and we will need to solve Eq.~\eqref{eq:wall_ap} numerically. We can simplify the integration upper limit by introducing $x = \alpha/s\beta_0$:
\begin{equation}
    \psi_w(\vett{r},t) = -2sa_1 \beta_0 e^{i(\beta_0 \eta + m\varphi)} \int_0^1 \mathrm{d}x \, x \, J_m(\omega_0 x) \, e^{-x^2/\Delta x^2},
\end{equation}
where  $\omega_0 = 2\rho\beta_0\sqrt{s}$ and $\Delta x^2 = 1/s\beta_0(a_1+i\xi)$.

\subsection*{Splash Pulses}
Introducing the splash spectrum [Eq.~\eqref{eq:splashspectrum}] into the expression for the generalized solution of the scalar wave equation inside the fiber core [Eq.~\eqref{eq:core}] and integrating first over $\kappa_1$ and then over $\beta$, we get
\begin{subequations}
\begin{equation}
    \psi_u(\vett{r},t) = \frac{a_1a_2}{2} \, \gamma \, e^{im\varphi} \, J_m(\gamma \rho) \int_0^\infty \frac{\mathrm{d}\alpha}{\alpha} \, \exp\left[-\alpha(a_1+i\xi) - \frac{\gamma^2}{4\alpha}(a_2-i\eta)\right],
\end{equation}
\begin{equation}
    \psi_w(\vett{r},t) = - \frac{a_1a_2}{2} \, \gamma \, e^{im\varphi} \, J_m(\gamma \rho) \int_0^{\gamma\sqrt{s}/2} \frac{\mathrm{d}\alpha}{\alpha} \, \exp\left[-\alpha(a_1+i\xi) - \frac{\gamma^2}{4\alpha}(a_2-i\eta)\right].
\end{equation}
\end{subequations}

A closed-form solution for the unperturbed term can be found by using identity (3.471.12) from Gradshteyn and Ryzhik \cite{gradshteyn_table_2007}:
\begin{equation}
    \psi_u(\vett{r},t) = a_1 a_2 \gamma \, e^{im\varphi} \, J_m(\gamma \rho) \, K_0\left[\gamma \sqrt{(a_1+i\xi)(a_2-i\eta)}\right].
\end{equation}

The wall term has no closed form but we can simplify the integration upper limit with the change of variables $x = 2\alpha/\gamma\sqrt{s}$:
\begin{equation}
    \psi_w(\vett{r},t) = - a_1 a_2 \gamma \, e^{im\varphi} J_m(\gamma\rho) \int_0^1 \frac{\mathrm{d}x}{x} \exp\left[-\frac{\gamma\sqrt{s}}{2}(a_1+i\xi)x -\frac{\gamma}{2x\sqrt{s}}(a_2-i\eta)\right].
\end{equation}  

\section*{References}
\bibliographystyle{unsrt} 
\bibliography{citations}

\end{document}